\documentclass[aps, reprint, showpacs]{revtex4-1} 
\usepackage{amsmath,amssymb,amsfonts, graphics,graphicx,color,multirow}

\newcommand{\rmd}{\mathrm{d}}

\newcommand{\bA}{\mathbf{A}}

\newcommand{\dxdt}{\dot{x}}
\newcommand{\ddxdt}{\ddot{x}}

\newcommand{\tf}{\tilde{f}}
\newcommand{\tx}{\tilde{x}}

\newcommand{\bchi}{\boldsymbol\chi}
\newcommand{\bkappa}{\boldsymbol\kappa}
\newcommand{\bb}{\mathbf{b}}
\begin{document}
\title{Energy transfer in networks with local magnetic time-reversal symmetry breaking}
\author{Benedikt \surname{Sabass}}
\affiliation{MAE, Princeton University, Princeton, New Jersey 08544, USA}
\email{bsabass@princeton.edu}
\begin{abstract}
Time-reversal symmetry of most conservative forces constrains the properties of linear transport in most physical systems. Here, I study the efficiency of energy transfer in oscillator networks where time-reversal symmetry is broken locally by Lorentz-force-like couplings. Despite their linearity, such networks can exhibit mono-directional transport and allow to isolate energy transfer in subsystems. New mechanisms and general rules for mono-directional transport are discussed. It is shown that the efficiency at maximum power can exceed $1/2$ and may even approach the upper bound of unity. 
\end{abstract}
\pacs{05.60.Cd, 05.45.Xt, 05.70.Ln}
%05.70.Ln 	Nonequilibrium and irreversible thermodynamics
%05.60.Cd 	Classical transport
%05.45.Xt coupled oscillators
\maketitle
\paragraph{Introduction.--}Onsager symmetry holds for the vast majority of coupled linear phenomena on the mesoscopic 
and macroscopic scale. However, the symmetry of transport coefficients is generally lost when time-reversal symmetry is broken through a magnetic field. Recently, it has been suggested that magnetic breaking of Onsager symmetry may allow for the existence of finite-time heat engines with vanishing entropy production and also allow an efficiency at maximum power that exceeds the Curzon-Ahlborn limit~\cite{BenentiSaitoCasati2011}. Although following studies established a number of positive lower bounds for entropy production in concrete quantum-mechanical systems~\cite{buttiker1985generalized,PhysRevB.85.085401,BrandnerSeifert2013,WhitneyQuantumThermoel}, and classical heat engines~\cite{AllahverdyanPRL2013, StarkSeifert2014}, the results highlight the non-trivial nature of coupled transport in the presence of a magnetic field. A logical next step is to combine the intriguing properties of magnetic couplings with topological features of a network. 

\begin{figure}
  \centering
  \includegraphics[width=\linewidth]{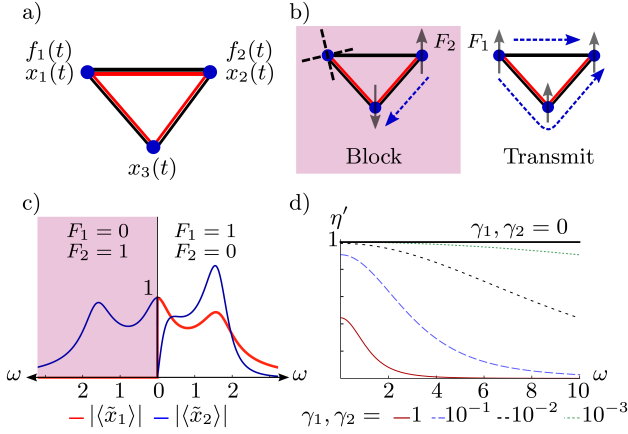} %[scale=1]
\caption{
\textbf{a)}~three-oscillator system with variables $x_{1,2,3}$ where oscillators $1$ and $2$ can be forced. Red lines symbolize magnetic couplings. \textbf{b)}~working principle of the three-oscillator diode. \textbf{c)} exemplary diode response ($\gamma_1=0.2$, $\gamma_2=0.1$). Left:~blocking state, Right:~transmitting state. \textbf{d)}~efficiency at maximum power $\eta'$ of the diode approaches unity when $\gamma_1$ and $\gamma_2$ vanish ($F_1=1$). \textbf{Parameters:}~$\kappa_{11}=\kappa_{22}=2$, $\kappa_{13}=\kappa_{23}=-1$, $b_{23}=1$, $\gamma_3 =2$.} 
\label{fig_diode_example}
\end{figure}

Networks with classical oscillators, such as vibrating spring-damper collections, electric power grids~\cite{RhodenPowerGrid}, or electronic circuits are omnipresent in our daily life. Desired system properties, e.g., in terms of frequency response or current rectification, often require active or non-linear elements, such as amplifiers or diodes. However, linear elements that break Onsager symmetry could be a possible alternative. Such elements can be built in a variety of ways: Mechanically, a coupling via Lorentz forces can be realized with a charge on a forced, two-dimensional pendulum in a magnetic field. Electrical circuits can be connected via perpendicular sides of a Hall element, which is called a gyrator~\cite{tellegen1948gyrator,grubbs1959hall}. In the microwave domain, similar devices employ the Faraday effect~\cite{RevModPhys.25.253}. In principle, even Coriolis forces can be employed. Here, such Lorentz-force-like couplings are combined into networks. Although the studied systems are restricted enough to be generic, they exhibit a surprisingly rich linear response and unusual energetics.
\paragraph{Framework.--}Our networks consist of coupled real variables $x_i(t)$ that obey a Langevin equation as
\begin{align}
\ddxdt_i &= -\sum_j\left(\kappa_{ij}\,x_j + b_{ij}\,\dxdt_j\right)-\gamma_i\,\dxdt_i + \xi_i + f_i.
\label{eq_network}
\end{align}
The symmetric matrix $\bkappa=\bkappa^{T}$ represents, e.g., elastic constants for a mechanical system or capacitance for an electric network. $\bkappa$ must be positive semi-definite for stability~\cite{strogatz2006nonlinear}. The antisymmetric matrix $\bb = -\bb^{T}$ represents Lorenz-force-like couplings. $\gamma_i$ are friction constants. The noise $\xi_i$ obeys $\langle\xi_i(t)\rangle = 0$ and $\langle \xi_i(t)  \xi_j(t')\rangle = \delta_{ij} 2 \gamma_i k_{\rm b}T \delta(t-t')$ where $k_{\rm b}T$ is the thermal energy. Only the two oscillators with indices $a$ and $b$ are driven as
\begin{align}
f_{\{a,b\}} = 2\,F_{\{a,b\}} \cos(\omega t + \varphi_{\{a,b\}} ) , & & f_{i\,\notin \{a,b\} } = 0, 
\end{align}
where $\omega$ is a fixed angular frequency. Phase differences will be written as $\varphi_{ij}\equiv\varphi_{i}-\varphi_{j}$. Non-dimensional units \cite{Nondim_note} are used throughout the letter. Variables in Fourier space will carry a tilde~$(\,\tilde{}\,)$.\\
The system's response to the driving forces is determined by the complex admittance 
\begin{align}
\bchi =\bA^{-1} \equiv \left[(-\omega^2+ i\,\omega\, \gamma_i )\delta_{ij}+\kappa_{ij}+i\,\omega\,b_{ij} \right]^{-1}. 
\label{eq_complex_admittance}
\end{align}
When $\bb = 0$, the conservative forces are time-reversal symmetric. Then, the complex admittance is symmetric $\bchi = \bchi^{T}$~\cite{Kubo1992statistical}, which is usually referred to as Onsager symmetry.\\
The energetics in steady state is governed by average work rates at the actuated oscillators 
$\dot{W}_{\{a,b\}} \equiv \frac{\omega}{2 \pi}\int_0^{\frac{2 \pi}{\omega}}  f_{\{a,b\}}(t)\,\langle\dxdt_{\{a,b\}}\rangle \,\rmd t$. The overall dissipation is given by the sum of the work rates as $\dot{W}_{\rm diss} \equiv \dot{W}_a+\dot{W}_b = \sum_j 2\,\gamma_j \omega^2 |\langle\tx_j\rangle|^2$. A net energy transmission through the system occurs when the power at one of the driven oscillators becomes negative.  For broken Onsager symmetry, the choice of input and output oscillators generally matters. In the following, the index $b$ will be used for the oscillator that provides energy output as $\dot{W}_{b}<0$. Then, the efficiency of energy transfer can be defined as
\begin{align}
\eta \equiv -\dot{W}_b/\dot{W}_a = -\dot{W}_b/(-\dot{W}_b + \dot{W}_{\rm diss}) \leq 1.
\label{eq_eff_definition}
\end{align}
A key figure of merit is efficiency at maximum power output, which is here defined for fixed frequencies as 
\begin{align}
\eta'  \equiv \eta(-\dot{W}_b |_{\substack{\omega}} \rightarrow \mathrm{max}),
\label{eq_effmaxp_def}
\end{align}
where the maximum $-\dot{W}'_b|_{\substack{\omega}}$ of the power output is found by searching for an optimal phase difference $\varphi'_{ba}$ and magnitude of the driving force $f'_b$.\\
The following three examples provide intuition on the interplay of network structure and local breaking of Onsager symmetry in systems of the type of Eq.~(\ref{eq_network}).
\paragraph{Example 1: A highly efficient diode.--}Consider a network of three oscillators as depicted in Fig.~\ref{fig_diode_example}a). The dynamics is described by Eq.~(\ref{eq_network}) where $j=1\ldots3$. Broken Onsager symmetry allows to choose the coupling parameters such as to have $\chi_{12}(\omega)=0$ for any frequency. This condition in Eq.~(\ref{eq_complex_admittance}) yields the parameters $b_{12} = 0,\; b_{13} = -\kappa_{12}/b_{23},\; \kappa_{33} = \kappa_{13} \kappa_{23}/\kappa_{12}$, and $\kappa_{12} = -(\kappa_{13} b_{23}^2)/(\kappa_{23} + b_{23} \gamma_3)$. Note that stability conditions on $\bkappa$ now impose a stronger constraint on the remaining free parameters. As illustrated in Fig.~\ref{fig_diode_example}b,c) the system is a dynamic, but genuine diode. Forcing with $f_2$ always leads to an anti-phase effect of oscillators $2$ and $3$ on oscillator $1$, which completely blocks the response of the latter. In contrast, forcing with $f_1$ leads to no cancellation of $1$ and $3$, such that energy can be carried around both sides of the structure. 

\begin{figure}
  \centering	
  \includegraphics[width=\linewidth]{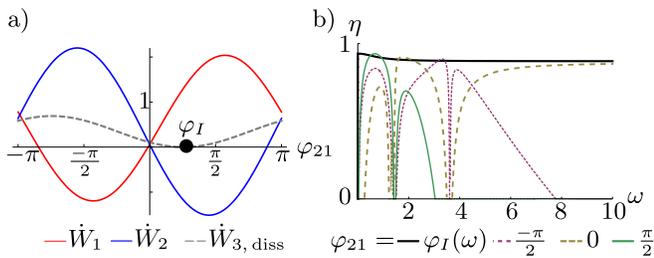}
\caption{The three-oscillator system shown in Fig.~\ref{fig_diode_example}a) can exhibit isolated energy transfer. \textbf{a)}~work rates $\dot{W}_{1,2}$ and dissipation in oscillator $3$ defined as \small{$\dot{W}_{3,\,\rm{diss}}\equiv \frac{\omega}{2 \pi} \int_0^{\frac{2 \pi}{\omega}} \gamma_3\langle \dot{x}_3 (t)\rangle^2\,\rmd t$}. Isolation occurs at a frequency-dependent phase $\varphi_I(\omega)$ (here $\omega=1.25$). \textbf{b)}~efficiency $\eta(\omega)$ at $\varphi_I(\omega)$ and at fixed values of $\varphi_{21}$. \textbf{Parameters:} $\kappa_{ii}=2$, $\kappa_{i\neq j}=-1$, $b_{12}=b_{23}=-2$, $b_{13}=-1$, $\gamma_{1,2,3} =0.1$, $F_1=1$, $|f_2(\omega)|$ is chosen according to Eq.~(\ref{eq_isolation}).} 
\label{fig_isolation_example}
\end{figure}

Fig.~\ref{fig_diode_example}d) shows the efficiency at maximum power $\eta'$. Generally, $\eta'$ has here only one maximum located at $\omega=0$. As shown, the system can asymptotically reach a unit efficiency at maximum power. This upper bound is achieved when friction in oscillators $1$ and $2$ vanish since $\lim_{(\gamma_1,\gamma_2) \rightarrow 0}\eta' = 1$. In contrast to systems with conserved Onsager symmetry, this diode can reach its upper bound of efficiency at almost arbitrary values of $\gamma_3$. For $\gamma_3 \rightarrow \infty$, the upper bound of $\eta'$ is reached when $\gamma_{1,2}$ tend to zero as $\gamma_3^{-h}$ with $h>1$. In the limit $\gamma_3 \rightarrow 0$, the system always becomes unstable since $\bkappa$ is then no longer positive definite. It should be emphasized that the theoretical reachability of $\eta'\approx 1$ in a genuine linear-response steady-state makes this new diode truly remarkable.
\paragraph{Example 2: Isolation in a three-oscillator network.--}The network in Fig.~\ref{fig_diode_example}a) can also be used to demonstrate how Lorentz-force-like couplings allow localization of energy transfer. If $\langle \dot{x}_3(t)\rangle$ vanishes during energy transfer between oscillators $1$ and $2$, no losses occur in oscillator $3$ and the state is called isolated. Here, this state is not achieved by tuning the system parameters, but by choosing proper driving forces $f_1(t)$, $f_2(t)$. Fig.~\ref{fig_isolation_example}a) exemplifies the occurrence of isolated energy transfer for a fixed phase lag between the forces.\\
A qualitative understanding can be gained by noting that the Fourier coefficients obey in the isolated state 
\begin{align}  
c\equiv \langle\tx_1\rangle/\langle\tx_2\rangle = -(\kappa_{23} - i b_{23} \omega)/(\kappa_{13} - i b_{13} \omega).
\label{eq_xAtIsolation}
\end{align}
Eq.~(\ref{eq_xAtIsolation}) determines the relative phase $\phi_{12}$ beween $x_1(t)$ and $x_2(t)$. $\phi_{12}$ in turn affects the energy transfer between the actuated oscillators. For usual networks with conserved Onsager symmetry, Eq.~(\ref{eq_xAtIsolation}) requires $\phi_{12} \in \{n\pi,\,n \in \mathbb{Z}\}$, while nonzero energy transfer would then require the opposite condition $\phi_{12} \notin \{n\pi,\,n \in \mathbb{Z}\}$. Thus, energy transfer would always lead to dissipation in the third oscillator if Onsager symmetry were conserved.  

\begin{figure}
  \centering
  \includegraphics[width=\linewidth]{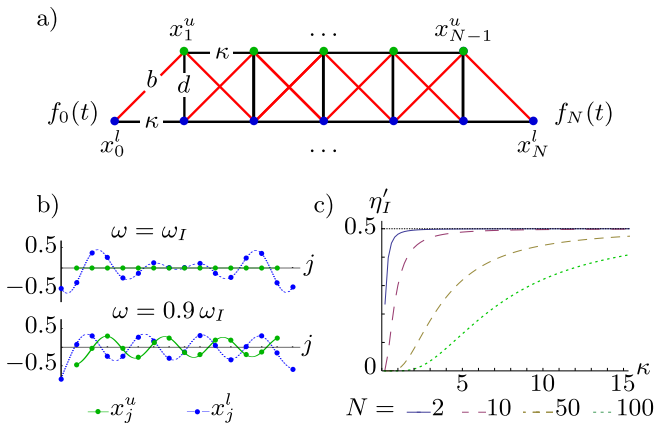}
\caption{
\textbf{a)}~oscillator system that allows mono-directional isolation of the upper chain from energy transfer in the lower chain. \textbf{b)}~exemplary responses of the chains driven at isolation frequency $\omega_I$ and below ($t=0$, $N= 15$). \textbf{c)}~efficiency at maximum power for isolated energy transfer through the chain. The effect of the magnetic coupling vanishes for the lower chain in the isolated state, thus $\eta'_I\leq1/2$.  \textbf{Parameters:}~$F_0=1$, $d=0.1$,  $\gamma_u=0.1$.} 
\label{fig_chain_example}
\end{figure}

\begin{figure}[htb]
  \centering
  \includegraphics[width=\linewidth]{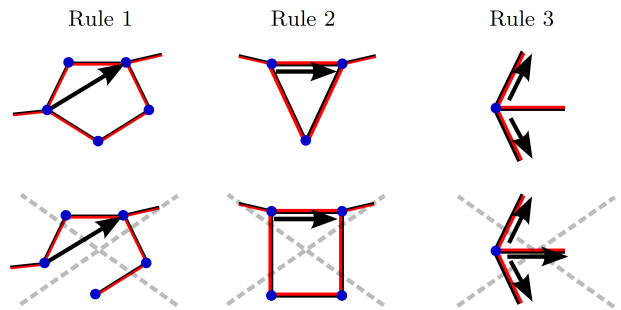}
\caption{Illustration of rules for mono-directional transport. Black arrows are the mono-directional links. Red/Black lines represent any kind of coupling with $b_{ij}$ or $\kappa_{ij}$. The three upper networks fulfill the rules while the corresponding networks below are impossible.} 
\label{fig_Rules}
\end{figure}

In this example, efficiency at maximum power $\eta'$ (Eq.~(\ref{eq_effmaxp_def})), is a useless concept since the forces are already fixed. The full efficiency in the isolated state $\eta_I$ can be written conveniently in Fourier space by denoting imaginary and real parts with $\Im()$ and $\Re()$. The result is
\begin{equation}
\eta_I = -(\Im(A_{22}) + \Im(c A_{21}))/(c^{*} c^{}\Im(A_{11}) + \Im(c^{*} A_{12}^{})),
\label{eq_eta_I}
\end{equation}
which is defined for $\dot{W}_2 \leq 0$. Eq.~(\ref{eq_eta_I}) results from insertion of Eq.~(\ref{eq_xAtIsolation}) into Eq.~(\ref{eq_eff_definition}). Remarkably, $\eta_I$ has a very simple frequency-dependence since there is only one extremum, located at $\omega=0$. For high $\omega$, $\eta_I$ can approach a non-zero constant; see Eq.~(\ref{eq_eta_high_omega_limit}) below. The conditions $b_{13} = b_{23},\,\kappa_{13} = \kappa_{23}$ even allow a frequency independent efficiency as $\eta_I = (b_{12} -\gamma_1)/(b_{12} +\gamma_2)$. Fig.~\ref{fig_isolation_example}b) illustrates that $\eta_I$ is not always the maximum efficiency since resonances in states with no isolation can lead to localized efficiency peaks. Thus, isolation with magnetic coupling does not necessarily lead to efficiency records, but can instead offer high values of $\eta_I$ throughout the whole frequency spectrum.
\paragraph{Example 3: Isolated transmission chain.--}The working principle of the diode can be employed to 
allow for isolation of energy transfer in spatially extended systems. Consider two chains as shown in Fig~\ref{fig_chain_example}a), consisting of $N+1$ and $N-1$ oscillators respectively. The oscillating variables of upper and lower chain $x^{u}_{j}$, $x^{l}_{j}$ obey linear equations~\cite{ChainBcs_note}, where noise is dropped for simplicity. The sought-for isolation mechanism should hold for waves travelling in both directions. Therefore, the system is made left-right symmetric by giving the two magnetic couplings at each oscillator always the same sign. The wave modes $(\tx^{u}_{\alpha},\tx^{l}_{\alpha})\,e^{i(\omega t -  k_{\alpha}\,j)}$ satisfy
\begin{small}
\begin{align}
\begin{split}
&\begin{pmatrix}
 w^2-2\kappa(1-\cos(k_{\alpha}))-d-i\omega \gamma_u \;\;\,\,d+i\omega 2b\cos(k_{\alpha})  \\
 d-i\omega 2b \cos(k_{\alpha})\;\;\,\,w^2-2\kappa(1-\cos(k_{\alpha}))-d-i\omega \gamma_l
\end{pmatrix}
\begin{pmatrix}
\tx^{u}_{\alpha}\\
\tx^{l}_{\alpha}
\end{pmatrix}=0.
\label{eq_chain_bulk_2}
\end{split}
\end{align} 
\end{small}
Asymmetry of the matrix in Eq.~(\ref{eq_chain_bulk_2}) allows $\tx^{u}$ to be independent of $\tx^{l}$ if the upper off-diagonal and the lower diagonal entries vanish. Such a state can be achieved if the parameters are chosen as $b = \kappa/\sqrt{d + 2\kappa}$ and $\gamma_{l} = d/\sqrt{d + 2\kappa}$. On using a forcing frequency $\omega_I\equiv\sqrt{d + 2\kappa}$, Eq.~(\ref{eq_chain_bulk_2}) yields the wave vectors $k_{1,2} =\pm\arccos(i\,d/(2\kappa))+2\pi\,n$, $n\in \mathbb{Z}$. Given proper boundary conditions, we then have a one-way isolation of the upper chain from the lower chain as illustrated by Fig~\ref{fig_chain_example}b). Note that isolation is here independent of $\varphi_{0 N}$ and $F_{\{0,N\}}$. Energy transmission is now also insensitive to spatial variations or friction in the upper chain. For conserved Onsager symmetry, only a two-way isolation is possible, albeit very hard to achieve in spatially extended systems, due to dissipation.
\paragraph{Mono-directional transport in arbitrary networks.--}Diode-like directional links as in example 1 can emerge in any
network with broken Onsager symmetry when off-diagonal elements of the complex admittance vanish asymmetrically. If excitations are to travel from oscillator $i$ to $j$ but not the reverse way, the following conditions must hold for any $\omega$
\begin{align}
\chi_{ji} \neq 0, & & \chi_{ij} \sim \det\left(\bA_{(ji)}\right) = 0,
\label{eq_directed_condition}
\end{align}
where $\bA_{(ji)}$ is the submatrix of $\bA$ that results when row $j$ and column $i$ are eliminated. To design a network with these properties, system parameters must be determined by solving Eq.~(\ref{eq_directed_condition}) for all orders of $\omega$, which can be tedious. However, the network topology already provides certain clues about the possibility of mono-directional links in networks of type (\ref{eq_network}). The following general rules (Fig.~\ref{fig_Rules}) are seen to hold~\cite{sabassToBePublished}  
\begin{enumerate}
	\item Mono-directional links require network loops.
	\item A mono-directional link between immediately coupled oscillators requires a loop of exactly three oscillators.
	\item No oscillator can have only mono-directional links.
\end{enumerate}
Rule 3 also admits an interesting thermodynamic interpretation: Consider a force-free system where the oscillators are embedded in different heat baths. Heat exchange is given by the deviations of the kinetic oscillator temperatures from the temperatures $T_i$ of the heat baths as $\dot{Q}_i = \gamma_i(\langle \dot{x}_i^2\rangle - k_{\rm b}\,T_i)$ \cite{sekimoto1998langevin}. If it were possible to connect oscillator $j$ in a totally mono-directional way to the rest of the network, heat would always flow either towards $j$, or away from $j$, regardless of temperature difference. Thus, such a network would violate the second law. 

Note that a diode could also be made with only one magnetic coupling $b_{ab}$ if a similarly strong frictional / resistive coupling was present~\cite{mason1953hall,wick1954solution}. However, the latter causes extra losses, and thus is not considered here.
\paragraph{Isolated energy transfer in arbitrary networks.--}As example 2 demonstrates, isolated energy transfer can be achieved without tuning of the coupling parameters. Since the mechanism of isolation instead depends on forces, phases, and frequencies, it can be understood as a genuine interference effect. For a general network, isolation of the oscillators with indices $\{m\}$ is defined as having
\begin{align}
\sum_{j\notin\{m\}}\chi_{mj} \tf_j = 0,
\label{eq_isolation}
\end{align}
which amounts to $\langle x_{\{m\}}\rangle=0$ if no forces are present at $\{m\}$. Equivalently, $\sum_{i\notin\{m\}}A_{m i} \langle\tx_i\rangle = 0$ can be demanded. Interferent isolation does not always require broken Onsager symmetry. If oscillator $m$ is directly connected to three or more energy-transmitting oscillators, coupling via the symmetric matrix $\bkappa$ is sufficient. On the other hand, $\bb\neq0$ allows isolation with only two directly connected oscillators (example 2). Moreover, the phase-shifts caused by $\bb$ can strongly increase the size of the frequency window where energy transfer takes place.\\ 
Generally, isolation does not minimize dissipation since, speaking in loose terms, the convex function $\dot{W}_{\rm diss}$ is minimized when power is distributed to many oscillators, rather than concentrated on a few. Isolation by interference is therefore primarily useful when the excitation of certain degrees of freedom is not wanted, e.g., when the overall response is otherwise dominated by them.
\paragraph{General formulas for efficiency.--}The work rates can be calculated conveniently by making use of Parceval's theorem. 
With $i\in\{a,b\}$, we have $\dot{W}_i  = -2\omega|\tf_i|^2 \Im(\chi_{ii})-2\omega \sum_{j\neq i} |\tf_i||\tf_j| \alpha_{ij}$,
where $\alpha_{ij}$ is a function of the phase difference $\varphi_{ij}$ as $\alpha_{ij} \equiv \Im(\chi_{ij})\cos(\varphi_{ij})-\Re(\chi_{ij})\sin(\varphi_{ij})$. At maximum power output $\dot{W}'_b$ we have $\tan{\varphi'_{ba}} = -\Re{(\chi_{ba})}/\Im{(\chi_{ba})}$ and $|\tf'_a|/|\tf'_b| = -2 \Im{(\chi_{bb})}/{\alpha'_{ba}}$ with $\alpha'_{ba}=\alpha_{ba}(\varphi'_{ba})$. Some care must be taken when selecting $\varphi'_{ba}$ since work is periodic in the phase. The efficiency at maximum power results to
\begin{align}
\eta' = \frac{{\alpha'}^2_{ba}}{2 \left( 2\Im(\chi_{bb})\Im(\chi_{aa}) - \alpha'_{ba} \alpha'_{ab}\right)} \leq 1.
\label{eq_effmaxp}
\end{align}
As demonstrated for the diode in example 1, the bound $\eta'=1$ can indeed be saturated asymptotically when Onsager symmetry is broken.\\
On the other hand, Onsager symmetry leads to a stronger bound when $\bb=0$ holds. We then have $\alpha'_{ab} = -\alpha'_{ba}(\cos^2(\varphi'_{ba})-\sin^2(\varphi'_{ba}))$. The efficiency at maximum power for any such network becomes
\begin{equation}
\eta'_s = \frac{1}{2\left(\frac{2\Im(\chi_{bb})\Im(\chi_{aa}) \cos^2(\varphi'_{ba})}{\Im(\chi_{ba})^2}  - \cos^2(\varphi'_{ba}) + \sin^2(\varphi'_{ba})\right)} \leq \frac{1}{2}.
\label{eq_eff_maxp_CA}
\end{equation}
The last inequality follows from $\Im(\chi_{bb})\Im(\chi_{aa})-\Im(\chi_{ab})^2 \geq 0$, which is for symmetric $\bchi$ equivalent to $\dot{W}_{\rm diss} \geq 0$. The bound $\eta'_s \leq 1/2$ is analogous to the Curzon-Ahlborn limit for heat machines~\cite{curzon1975efficiency,PhysRevLett.102.130602,PhysRevLett.108.210602}.

A further energetic advantage of magnetic coupling occurs when a direct link $b_{ab}\neq 0$ exists between input and output.
Then, energy transmission in the limit $\omega \rightarrow \infty$ scales with the same power of $\omega$ as frictional dissipation. This feature allows for tunable efficiency in high-frequency regimes. All systems of type (\ref{eq_network}) have the same high-frequency limit for efficiency 
\begin{align}
\lim\limits_{\omega \rightarrow \infty} \eta = \mathrm{max}\left(\frac{-|\tf_b|^2 \gamma_b + |\tf_a| |\tf_b| b_{ab} \cos{\varphi_{ab}}}{|\tf_a|^2 \gamma_a + |\tf_a||\tf_b| b_{ab} \cos{\varphi_{ab}}},0\right).
\label{eq_eta_high_omega_limit}
\end{align} 
\paragraph{Fluctuations.--}The additive noise in Eq.~(\ref{eq_network}) neither affects
average energy transfer nor the mechanism of directional transport. Multiplicative noise that
results from parameter fluctuations, in particular from unstable magnetic fields, could have 
a more pronounced effect. However, as shown in the supplement, white noise in $\bb$ on average 
merely renormalizes the friction constants, which does not affect the key principles described in this letter.
\paragraph{Concluding remarks.--}Although linear oscillators are a paradigm of established physics, the energetics of oscillator networks with Lorentz-force-like couplings have hardly been explored. Focussing on networks with only the most generic types of coupling, 
it has been shown here that unusual transport properties entail favorable energetics and result from the interplay of network topology and Onsager symmetry breaking.\\
The studied systems can be extended and used in many ways. They may be complemented with non-linear oscillators, e.g., to study synchronization phenomena~\cite{julicher2009spontaneous,martens2013chimera,JuelicherPhaseDelayPRL}. Periodic structures related to the oscillator chain in example 3 can allow for cloaking~\cite{Pendry23062006,Wegener2013metamaterials} -inspired mono-directional shielding~\cite{sabassToBePublished}. Mechanical applications, such as directional vibration damping, are at least in principle conceivable. Realization of the suggested high efficiency in electric circuits hinges on the availability of low-resistance symmetry-breaking couplings~\cite{wick1954solution, zhai2006quasi, PhysRevX.4.021019}. With these, the studied networks would become magnetic-field programmable and operate much like a transistor network, with the major distinction of being linear.
\begin{acknowledgements}
The author thanks Prof.~U.~Seifert for long-standing support and for discussions about the manuscript. A postdoctoral fellowship from the DAAD is acknowledged.
\end{acknowledgements}
%\bibliography{Energy_transmission_with_B_field_bib}

\begin{thebibliography}{29}%
\makeatletter
\providecommand \@ifxundefined [1]{%
 \@ifx{#1\undefined}
}%
\providecommand \@ifnum [1]{%
 \ifnum #1\expandafter \@firstoftwo
 \else \expandafter \@secondoftwo
 \fi
}%
\providecommand \@ifx [1]{%
 \ifx #1\expandafter \@firstoftwo
 \else \expandafter \@secondoftwo
 \fi
}%
\providecommand \natexlab [1]{#1}%
\providecommand \enquote  [1]{``#1''}%
\providecommand \bibnamefont  [1]{#1}%
\providecommand \bibfnamefont [1]{#1}%
\providecommand \citenamefont [1]{#1}%
\providecommand \href@noop [0]{\@secondoftwo}%
\providecommand \href [0]{\begingroup \@sanitize@url \@href}%
\providecommand \@href[1]{\@@startlink{#1}\@@href}%
\providecommand \@@href[1]{\endgroup#1\@@endlink}%
\providecommand \@sanitize@url [0]{\catcode `\\12\catcode `\$12\catcode
  `\&12\catcode `\#12\catcode `\^12\catcode `\_12\catcode `\%12\relax}%
\providecommand \@@startlink[1]{}%
\providecommand \@@endlink[0]{}%
\providecommand \url  [0]{\begingroup\@sanitize@url \@url }%
\providecommand \@url [1]{\endgroup\@href {#1}{\urlprefix }}%
\providecommand \urlprefix  [0]{URL }%
\providecommand \Eprint [0]{\href }%
\providecommand \doibase [0]{http://dx.doi.org/}%
\providecommand \selectlanguage [0]{\@gobble}%
\providecommand \bibinfo  [0]{\@secondoftwo}%
\providecommand \bibfield  [0]{\@secondoftwo}%
\providecommand \translation [1]{[#1]}%
\providecommand \BibitemOpen [0]{}%
\providecommand \bibitemStop [0]{}%
\providecommand \bibitemNoStop [0]{.\EOS\space}%
\providecommand \EOS [0]{\spacefactor3000\relax}%
\providecommand \BibitemShut  [1]{\csname bibitem#1\endcsname}%
\let\auto@bib@innerbib\@empty
%</preamble>
\bibitem [{\citenamefont {Benenti}\ \emph {et~al.}(2011)\citenamefont
  {Benenti}, \citenamefont {Saito},\ and\ \citenamefont
  {Casati}}]{BenentiSaitoCasati2011}%
  \BibitemOpen
  \bibfield  {author} {\bibinfo {author} {\bibfnamefont {G.}~\bibnamefont
  {Benenti}}, \bibinfo {author} {\bibfnamefont {K.}~\bibnamefont {Saito}}, \
  and\ \bibinfo {author} {\bibfnamefont {G.}~\bibnamefont {Casati}},\ }\href
  {\doibase 10.1103/PhysRevLett.106.230602} {\bibfield  {journal} {\bibinfo
  {journal} {Phys. Rev. Lett.}\ }\textbf {\bibinfo {volume} {106}},\ \bibinfo
  {pages} {230602} (\bibinfo {year} {2011})}\BibitemShut {NoStop}%
\bibitem [{\citenamefont {B{\"u}ttiker}\ \emph {et~al.}(1985)\citenamefont
  {B{\"u}ttiker}, \citenamefont {Imry}, \citenamefont {Landauer},\ and\
  \citenamefont {Pinhas}}]{buttiker1985generalized}%
  \BibitemOpen
  \bibfield  {author} {\bibinfo {author} {\bibfnamefont {M.}~\bibnamefont
  {B{\"u}ttiker}}, \bibinfo {author} {\bibfnamefont {Y.}~\bibnamefont {Imry}},
  \bibinfo {author} {\bibfnamefont {R.}~\bibnamefont {Landauer}}, \ and\
  \bibinfo {author} {\bibfnamefont {S.}~\bibnamefont {Pinhas}},\ }\href@noop {}
  {\bibfield  {journal} {\bibinfo  {journal} {Phys. Rev. B}\ }\textbf {\bibinfo
  {volume} {31}},\ \bibinfo {pages} {6207} (\bibinfo {year}
  {1985})}\BibitemShut {NoStop}%
\bibitem [{\citenamefont {Entin-Wohlman}\ and\ \citenamefont
  {Aharony}(2012)}]{PhysRevB.85.085401}%
  \BibitemOpen
  \bibfield  {author} {\bibinfo {author} {\bibfnamefont {O.}~\bibnamefont
  {Entin-Wohlman}}\ and\ \bibinfo {author} {\bibfnamefont {A.}~\bibnamefont
  {Aharony}},\ }\href {\doibase 10.1103/PhysRevB.85.085401} {\bibfield
  {journal} {\bibinfo  {journal} {Phys. Rev. B}\ }\textbf {\bibinfo {volume}
  {85}},\ \bibinfo {pages} {085401} (\bibinfo {year} {2012})}\BibitemShut
  {NoStop}%
\bibitem [{\citenamefont {Brandner}\ \emph {et~al.}(2013)\citenamefont
  {Brandner}, \citenamefont {Saito},\ and\ \citenamefont
  {Seifert}}]{BrandnerSeifert2013}%
  \BibitemOpen
  \bibfield  {author} {\bibinfo {author} {\bibfnamefont {K.}~\bibnamefont
  {Brandner}}, \bibinfo {author} {\bibfnamefont {K.}~\bibnamefont {Saito}}, \
  and\ \bibinfo {author} {\bibfnamefont {U.}~\bibnamefont {Seifert}},\ }\href
  {\doibase 10.1103/PhysRevLett.110.070603} {\bibfield  {journal} {\bibinfo
  {journal} {Phys. Rev. Lett.}\ }\textbf {\bibinfo {volume} {110}},\ \bibinfo
  {pages} {070603} (\bibinfo {year} {2013})}\BibitemShut {NoStop}%
\bibitem [{\citenamefont {Whitney}(2014)}]{WhitneyQuantumThermoel}%
  \BibitemOpen
  \bibfield  {author} {\bibinfo {author} {\bibfnamefont {R.~S.}\ \bibnamefont
  {Whitney}},\ }\href {\doibase 10.1103/PhysRevLett.112.130601} {\bibfield
  {journal} {\bibinfo  {journal} {Phys. Rev. Lett.}\ }\textbf {\bibinfo
  {volume} {112}},\ \bibinfo {pages} {130601} (\bibinfo {year}
  {2014})}\BibitemShut {NoStop}%
\bibitem [{\citenamefont {Allahverdyan}\ \emph {et~al.}(2013)\citenamefont
  {Allahverdyan}, \citenamefont {Hovhannisyan}, \citenamefont {Melkikh},\ and\
  \citenamefont {Gevorkian}}]{AllahverdyanPRL2013}%
  \BibitemOpen
  \bibfield  {author} {\bibinfo {author} {\bibfnamefont {A.~E.}\ \bibnamefont
  {Allahverdyan}}, \bibinfo {author} {\bibfnamefont {K.~V.}\ \bibnamefont
  {Hovhannisyan}}, \bibinfo {author} {\bibfnamefont {A.~V.}\ \bibnamefont
  {Melkikh}}, \ and\ \bibinfo {author} {\bibfnamefont {S.~G.}\ \bibnamefont
  {Gevorkian}},\ }\href {\doibase 10.1103/PhysRevLett.111.050601} {\bibfield
  {journal} {\bibinfo  {journal} {Phys. Rev. Lett.}\ }\textbf {\bibinfo
  {volume} {111}},\ \bibinfo {pages} {050601} (\bibinfo {year}
  {2013})}\BibitemShut {NoStop}%
\bibitem [{\citenamefont {Stark}\ \emph {et~al.}(2014)\citenamefont {Stark},
  \citenamefont {Brandner}, \citenamefont {Saito},\ and\ \citenamefont
  {Seifert}}]{StarkSeifert2014}%
  \BibitemOpen
  \bibfield  {author} {\bibinfo {author} {\bibfnamefont {J.}~\bibnamefont
  {Stark}}, \bibinfo {author} {\bibfnamefont {K.}~\bibnamefont {Brandner}},
  \bibinfo {author} {\bibfnamefont {K.}~\bibnamefont {Saito}}, \ and\ \bibinfo
  {author} {\bibfnamefont {U.}~\bibnamefont {Seifert}},\ }\href {\doibase
  10.1103/PhysRevLett.112.140601} {\bibfield  {journal} {\bibinfo  {journal}
  {Phys. Rev. Lett.}\ }\textbf {\bibinfo {volume} {112}},\ \bibinfo {pages}
  {140601} (\bibinfo {year} {2014})}\BibitemShut {NoStop}%
\bibitem [{\citenamefont {Rohden}\ \emph {et~al.}(2012)\citenamefont {Rohden},
  \citenamefont {Sorge}, \citenamefont {Timme},\ and\ \citenamefont
  {Witthaut}}]{RhodenPowerGrid}%
  \BibitemOpen
  \bibfield  {author} {\bibinfo {author} {\bibfnamefont {M.}~\bibnamefont
  {Rohden}}, \bibinfo {author} {\bibfnamefont {A.}~\bibnamefont {Sorge}},
  \bibinfo {author} {\bibfnamefont {M.}~\bibnamefont {Timme}}, \ and\ \bibinfo
  {author} {\bibfnamefont {D.}~\bibnamefont {Witthaut}},\ }\href {\doibase
  10.1103/PhysRevLett.109.064101} {\bibfield  {journal} {\bibinfo  {journal}
  {Phys. Rev. Lett.}\ }\textbf {\bibinfo {volume} {109}},\ \bibinfo {pages}
  {064101} (\bibinfo {year} {2012})}\BibitemShut {NoStop}%
\bibitem [{\citenamefont {Tellegen}(1948)}]{tellegen1948gyrator}%
  \BibitemOpen
  \bibfield  {author} {\bibinfo {author} {\bibfnamefont {B.~D.}\ \bibnamefont
  {Tellegen}},\ }\href@noop {} {\bibfield  {journal} {\bibinfo  {journal}
  {Philips Res. Rep.}\ }\textbf {\bibinfo {volume} {3}},\ \bibinfo {pages} {81}
  (\bibinfo {year} {1948})}\BibitemShut {NoStop}%
\bibitem [{\citenamefont {Grubbs}(1959)}]{grubbs1959hall}%
  \BibitemOpen
  \bibfield  {author} {\bibinfo {author} {\bibfnamefont {W.}~\bibnamefont
  {Grubbs}},\ }\href@noop {} {\bibfield  {journal} {\bibinfo  {journal} {Bell
  Syst. Tech. J.}\ }\textbf {\bibinfo {volume} {38}},\ \bibinfo {pages} {853}
  (\bibinfo {year} {1959})}\BibitemShut {NoStop}%
\bibitem [{\citenamefont {Hogan}(1953)}]{RevModPhys.25.253}%
  \BibitemOpen
  \bibfield  {author} {\bibinfo {author} {\bibfnamefont {C.~L.}\ \bibnamefont
  {Hogan}},\ }\href {\doibase 10.1103/RevModPhys.25.253} {\bibfield  {journal}
  {\bibinfo  {journal} {Rev. Mod. Phys.}\ }\textbf {\bibinfo {volume} {25}},\
  \bibinfo {pages} {253} (\bibinfo {year} {1953})}\BibitemShut {NoStop}%
\bibitem [{\citenamefont {Strogatz}(1994)}]{strogatz2006nonlinear}%
  \BibitemOpen
  \bibfield  {author} {\bibinfo {author} {\bibfnamefont {S.~H.}\ \bibnamefont
  {Strogatz}},\ }\href@noop {} {\emph {\bibinfo {title} {Nonlinear dynamics and
  chaos}}}\ (\bibinfo  {publisher} {Perseus Publishing},\ \bibinfo {year}
  {1994})\BibitemShut {NoStop}%
\bibitem [{Non()}]{Nondim_note}%
  \BibitemOpen
  \href@noop {} {}\bibinfo {note} {{We fix a force scale $\hat{F}$ and a
  typical coupling constant $\hat{\kappa}$. Time is non-dimensionalized by
  $\hat{\tau} \equiv 1/\sqrt{\hat{\kappa}}$. Since Eq.~(\ref{eq_network}) is
  non-dimensionalized by $\hat{F}$, $x$ is scaled by $\hat{x} \equiv
  \hat{F}/\hat{\kappa}$. Accordingly, $b_{ij}$ and $\gamma_{ij}$ are
  non-dimensionalized by $\sqrt{\hat{\kappa}}$. All energies are scaled by
  $\hat{F} \hat{x}$ and power with $\hat{F} \hat{x} /\hat{\tau}$ }}\BibitemShut
  {NoStop}%
\bibitem [{\citenamefont {Hashitsume}\ \emph {et~al.}(1992)\citenamefont
  {Hashitsume}, \citenamefont {Toda}, \citenamefont {Kubo},\ and\ \citenamefont
  {Sait{\=o}}}]{Kubo1992statistical}%
  \BibitemOpen
  \bibfield  {author} {\bibinfo {author} {\bibfnamefont {N.}~\bibnamefont
  {Hashitsume}}, \bibinfo {author} {\bibfnamefont {M.}~\bibnamefont {Toda}},
  \bibinfo {author} {\bibfnamefont {R.}~\bibnamefont {Kubo}}, \ and\ \bibinfo
  {author} {\bibfnamefont {N.}~\bibnamefont {Sait{\=o}}},\ }\href@noop {}
  {\emph {\bibinfo {title} {Statistical physics II: nonequilibrium statistical
  mechanics}}}\ (\bibinfo  {publisher} {Springer},\ \bibinfo {year}
  {1992})\BibitemShut {NoStop}%
\bibitem [{Cha()}]{ChainBcs_note}%
  \BibitemOpen
  \href@noop {} {}\bibinfo {note} {The dynamics of the chains in Example 3
  obey\\ $\ddot{x}^{u}_{j} =\kappa \left[x^{u}_{j-1}+x^{u}_{j+1}-2 x^{u}_{j}
  \right] + d \left[x^{l}_{j}-x^{u}_{j}\right] -\gamma_u \dot{x}^{u}_{j}$\\
  \hspace*{0.7cm}$+b\left[\dot{x}^{l}_{j-1}+\dot{x}^{l}_{j+1}\right]$,\\
  $\ddot{x}^{l}_{j} =\kappa\left[x^{l}_{j-1}+x^{l}_{j+1}-2 x^{l}_{j}\right] + d
  \left[x^{u}_{j}-x^{l}_{j}\right]-\gamma_l\dot{x}^{l}_{j}$ \\
  \hspace*{0.7cm}$-b\left[\dot{x}^{u}_{j-1}+\dot{x}^{u}_{j+1}\right]$.\\ The
  boundary conditions read\\ $\ddot{x}^{l}_0 =\kappa [x^{l}_1- x^{l}_0]
  -\gamma_l \dot{x}^{l}_0 - b \dot{x}^{u}_1 + f_0,$\\ $\ddot{x}^{l}_N =\kappa
  [x^{l}_{N-1}- x^{l}_N] -\gamma_l \dot{x}^{l}_N - b \dot{x}^{u}_{N-1} +
  f_N,$\\ $\ddot{x}^{u}_1 =\kappa [x^{u}_2- x^{u}_1] + d
  \left[x^{l}_1-x^{u}_1\right] -\gamma_u \dot{x}^{u}_1 +b
  [\dot{x}^{l}_0+\dot{x}^{l}_2],$\\ $\ddot{x}^{u}_{N-1} =\kappa [x^{u}_{N-2}-
  x^{u}_{N-1}] + d [x^{l}_{N-1}-x^{u}_{N-1}] -\gamma_u \dot{x}^{u}_{N-1}$\\
  \hspace*{0.7cm}$+b [\dot{x}^{l}_{N}+\dot{x}^{l}_{N-2}]$}\BibitemShut
  {NoStop}%
\bibitem [{\citenamefont {Sabass}()}]{sabassToBePublished}%
  \BibitemOpen
  \bibfield  {author} {\bibinfo {author} {\bibfnamefont {B.}~\bibnamefont
  {Sabass}},\ }\href@noop {} {\bibinfo  {journal} {to be published}\
  }\BibitemShut {NoStop}%
\bibitem [{\citenamefont {Sekimoto}(1998)}]{sekimoto1998langevin}%
  \BibitemOpen
\bibfield  {journal} {  }\bibfield  {author} {\bibinfo {author} {\bibfnamefont
  {K.}~\bibnamefont {Sekimoto}},\ }\href@noop {} {\bibfield  {journal}
  {\bibinfo  {journal} {Progr. Theor. Phys. Supplement}\ }\textbf {\bibinfo
  {volume} {130}},\ \bibinfo {pages} {17} (\bibinfo {year} {1998})}\BibitemShut
  {NoStop}%
\bibitem [{\citenamefont {Mason}\ \emph {et~al.}(1953)\citenamefont {Mason},
  \citenamefont {Hewitt},\ and\ \citenamefont {Wick}}]{mason1953hall}%
  \BibitemOpen
  \bibfield  {author} {\bibinfo {author} {\bibfnamefont {W.}~\bibnamefont
  {Mason}}, \bibinfo {author} {\bibfnamefont {W.}~\bibnamefont {Hewitt}}, \
  and\ \bibinfo {author} {\bibfnamefont {R.}~\bibnamefont {Wick}},\ }\href@noop
  {} {\bibfield  {journal} {\bibinfo  {journal} {J. Appl. Phys.}\ }\textbf
  {\bibinfo {volume} {24}},\ \bibinfo {pages} {166} (\bibinfo {year}
  {1953})}\BibitemShut {NoStop}%
\bibitem [{\citenamefont {Wick}(1954)}]{wick1954solution}%
  \BibitemOpen
  \bibfield  {author} {\bibinfo {author} {\bibfnamefont {R.~F.}\ \bibnamefont
  {Wick}},\ }\href@noop {} {\bibfield  {journal} {\bibinfo  {journal} {J. Appl.
  Phys.}\ }\textbf {\bibinfo {volume} {25}},\ \bibinfo {pages} {741} (\bibinfo
  {year} {1954})}\BibitemShut {NoStop}%
\bibitem [{\citenamefont {Curzon}\ and\ \citenamefont
  {Ahlborn}(1975)}]{curzon1975efficiency}%
  \BibitemOpen
  \bibfield  {author} {\bibinfo {author} {\bibfnamefont {F.}~\bibnamefont
  {Curzon}}\ and\ \bibinfo {author} {\bibfnamefont {B.}~\bibnamefont
  {Ahlborn}},\ }\href@noop {} {\bibfield  {journal} {\bibinfo  {journal} {Am.
  J. Phys.}\ }\textbf {\bibinfo {volume} {43}},\ \bibinfo {pages} {22}
  (\bibinfo {year} {1975})}\BibitemShut {NoStop}%
\bibitem [{\citenamefont {Esposito}\ \emph {et~al.}(2009)\citenamefont
  {Esposito}, \citenamefont {Lindenberg},\ and\ \citenamefont {Van~den
  Broeck}}]{PhysRevLett.102.130602}%
  \BibitemOpen
  \bibfield  {author} {\bibinfo {author} {\bibfnamefont {M.}~\bibnamefont
  {Esposito}}, \bibinfo {author} {\bibfnamefont {K.}~\bibnamefont
  {Lindenberg}}, \ and\ \bibinfo {author} {\bibfnamefont {C.}~\bibnamefont
  {Van~den Broeck}},\ }\href {\doibase 10.1103/PhysRevLett.102.130602}
  {\bibfield  {journal} {\bibinfo  {journal} {Phys. Rev. Lett.}\ }\textbf
  {\bibinfo {volume} {102}},\ \bibinfo {pages} {130602} (\bibinfo {year}
  {2009})}\BibitemShut {NoStop}%
\bibitem [{\citenamefont {Van~den Broeck}\ \emph {et~al.}(2012)\citenamefont
  {Van~den Broeck}, \citenamefont {Kumar},\ and\ \citenamefont
  {Lindenberg}}]{PhysRevLett.108.210602}%
  \BibitemOpen
  \bibfield  {author} {\bibinfo {author} {\bibfnamefont {C.}~\bibnamefont
  {Van~den Broeck}}, \bibinfo {author} {\bibfnamefont {N.}~\bibnamefont
  {Kumar}}, \ and\ \bibinfo {author} {\bibfnamefont {K.}~\bibnamefont
  {Lindenberg}},\ }\href {\doibase 10.1103/PhysRevLett.108.210602} {\bibfield
  {journal} {\bibinfo  {journal} {Phys. Rev. Lett.}\ }\textbf {\bibinfo
  {volume} {108}},\ \bibinfo {pages} {210602} (\bibinfo {year}
  {2012})}\BibitemShut {NoStop}%
\bibitem [{\citenamefont {J{\"u}licher}\ \emph {et~al.}(2009)\citenamefont
  {J{\"u}licher}, \citenamefont {Dierkes}, \citenamefont {Lindner},
  \citenamefont {Prost},\ and\ \citenamefont
  {Martin}}]{julicher2009spontaneous}%
  \BibitemOpen
  \bibfield  {author} {\bibinfo {author} {\bibfnamefont {F.}~\bibnamefont
  {J{\"u}licher}}, \bibinfo {author} {\bibfnamefont {K.}~\bibnamefont
  {Dierkes}}, \bibinfo {author} {\bibfnamefont {B.}~\bibnamefont {Lindner}},
  \bibinfo {author} {\bibfnamefont {J.}~\bibnamefont {Prost}}, \ and\ \bibinfo
  {author} {\bibfnamefont {P.}~\bibnamefont {Martin}},\ }\href@noop {}
  {\bibfield  {journal} {\bibinfo  {journal} {Eur. Phys. J. E}\ }\textbf
  {\bibinfo {volume} {29}},\ \bibinfo {pages} {449} (\bibinfo {year}
  {2009})}\BibitemShut {NoStop}%
\bibitem [{\citenamefont {Martens}\ \emph {et~al.}(2013)\citenamefont
  {Martens}, \citenamefont {Thutupalli}, \citenamefont {Fourri{\`e}re},\ and\
  \citenamefont {Hallatschek}}]{martens2013chimera}%
  \BibitemOpen
  \bibfield  {author} {\bibinfo {author} {\bibfnamefont {E.~A.}\ \bibnamefont
  {Martens}}, \bibinfo {author} {\bibfnamefont {S.}~\bibnamefont {Thutupalli}},
  \bibinfo {author} {\bibfnamefont {A.}~\bibnamefont {Fourri{\`e}re}}, \ and\
  \bibinfo {author} {\bibfnamefont {O.}~\bibnamefont {Hallatschek}},\
  }\href@noop {} {\bibfield  {journal} {\bibinfo  {journal} {PNAS}\ }\textbf
  {\bibinfo {volume} {110}},\ \bibinfo {pages} {10563} (\bibinfo {year}
  {2013})}\BibitemShut {NoStop}%
\bibitem [{\citenamefont {J\"org}\ \emph {et~al.}(2014)\citenamefont {J\"org},
  \citenamefont {Morelli}, \citenamefont {Ares},\ and\ \citenamefont
  {J\"ulicher}}]{JuelicherPhaseDelayPRL}%
  \BibitemOpen
  \bibfield  {author} {\bibinfo {author} {\bibfnamefont {D.~J.}\ \bibnamefont
  {J\"org}}, \bibinfo {author} {\bibfnamefont {L.~G.}\ \bibnamefont {Morelli}},
  \bibinfo {author} {\bibfnamefont {S.}~\bibnamefont {Ares}}, \ and\ \bibinfo
  {author} {\bibfnamefont {F.}~\bibnamefont {J\"ulicher}},\ }\href {\doibase
  10.1103/PhysRevLett.112.174101} {\bibfield  {journal} {\bibinfo  {journal}
  {Phys. Rev. Lett.}\ }\textbf {\bibinfo {volume} {112}},\ \bibinfo {pages}
  {174101} (\bibinfo {year} {2014})}\BibitemShut {NoStop}%
\bibitem [{\citenamefont {Pendry}\ \emph {et~al.}(2006)\citenamefont {Pendry},
  \citenamefont {Schurig},\ and\ \citenamefont {Smith}}]{Pendry23062006}%
  \BibitemOpen
  \bibfield  {author} {\bibinfo {author} {\bibfnamefont {J.~B.}\ \bibnamefont
  {Pendry}}, \bibinfo {author} {\bibfnamefont {D.}~\bibnamefont {Schurig}}, \
  and\ \bibinfo {author} {\bibfnamefont {D.~R.}\ \bibnamefont {Smith}},\ }\href
  {\doibase 10.1126/science.1125907} {\bibfield  {journal} {\bibinfo  {journal}
  {Science}\ }\textbf {\bibinfo {volume} {312}},\ \bibinfo {pages} {1780}
  (\bibinfo {year} {2006})}\BibitemShut {NoStop}%
\bibitem [{\citenamefont {Kadic}\ \emph {et~al.}(2013)\citenamefont {Kadic},
  \citenamefont {B{\"u}ckmann}, \citenamefont {Schittny},\ and\ \citenamefont
  {Wegener}}]{Wegener2013metamaterials}%
  \BibitemOpen
  \bibfield  {author} {\bibinfo {author} {\bibfnamefont {M.}~\bibnamefont
  {Kadic}}, \bibinfo {author} {\bibfnamefont {T.}~\bibnamefont {B{\"u}ckmann}},
  \bibinfo {author} {\bibfnamefont {R.}~\bibnamefont {Schittny}}, \ and\
  \bibinfo {author} {\bibfnamefont {M.}~\bibnamefont {Wegener}},\ }\href@noop
  {} {\bibfield  {journal} {\bibinfo  {journal} {Rep. Prog. Phys.}\ }\textbf
  {\bibinfo {volume} {76}},\ \bibinfo {pages} {126501} (\bibinfo {year}
  {2013})}\BibitemShut {NoStop}%
\bibitem [{\citenamefont {Zhai}\ \emph {et~al.}(2006)\citenamefont {Zhai},
  \citenamefont {Li}, \citenamefont {Dong}, \citenamefont {Viehland},\ and\
  \citenamefont {Bichurin}}]{zhai2006quasi}%
  \BibitemOpen
  \bibfield  {author} {\bibinfo {author} {\bibfnamefont {J.}~\bibnamefont
  {Zhai}}, \bibinfo {author} {\bibfnamefont {J.}~\bibnamefont {Li}}, \bibinfo
  {author} {\bibfnamefont {S.}~\bibnamefont {Dong}}, \bibinfo {author}
  {\bibfnamefont {D.}~\bibnamefont {Viehland}}, \ and\ \bibinfo {author}
  {\bibfnamefont {M.}~\bibnamefont {Bichurin}},\ }\href@noop {} {\bibfield
  {journal} {\bibinfo  {journal} {J. Appl. Phys.}\ }\textbf {\bibinfo {volume}
  {100}},\ \bibinfo {pages} {124509} (\bibinfo {year} {2006})}\BibitemShut
  {NoStop}%
\bibitem [{\citenamefont {Viola}\ and\ \citenamefont
  {DiVincenzo}(2014)}]{PhysRevX.4.021019}%
  \BibitemOpen
  \bibfield  {author} {\bibinfo {author} {\bibfnamefont {G.}~\bibnamefont
  {Viola}}\ and\ \bibinfo {author} {\bibfnamefont {D.~P.}\ \bibnamefont
  {DiVincenzo}},\ }\href {\doibase 10.1103/PhysRevX.4.021019} {\bibfield
  {journal} {\bibinfo  {journal} {Phys. Rev. X}\ }\textbf {\bibinfo {volume}
  {4}},\ \bibinfo {pages} {021019} (\bibinfo {year} {2014})}\BibitemShut
  {NoStop}%
\end{thebibliography}
%
\end{document}